\documentclass[12pt]{article}
\usepackage{epsf}
\usepackage{amssymb}
\usepackage[utf8]{inputenc}
\usepackage[english]{babel}

\begin{document}

\title{
{\bf High-energy central exclusive production of the lightest vacuum resonance related to the soft Pomeron}}
\author{A.A. Godizov\thanks{E-mail: anton.godizov@gmail.com}\\
{\small {\it A.A. Logunov Institute for High Energy Physics}},\\ {\small {\it NRC ``Kurchatov Institute'', 142281 Protvino, Russia}}}
\date{}
\maketitle

\begin{abstract}
A simple model based on Regge approach is proposed for description of the central exclusive production (CEP) of the light tensor glueball lying on the 
Regge trajectory of the soft Pomeron.
\end{abstract}

\section*{1. Introduction}

Reactions of diffractive CEP of light vacuum resonances in high-energy collisions of protons $p+p\to p+R+p$ (signs ``+'' denote 
rapidity gaps) are a valuable source of information on the nonperturbative aspects of strong interaction. At present, they are actively studied by both 
experimentalists \cite{sikora,berretti} and theorists \cite{fiore,szczurek}.
\begin{figure}[ht]
\begin{center}
\epsfxsize=7.7cm\epsfysize=7.7cm\epsffile{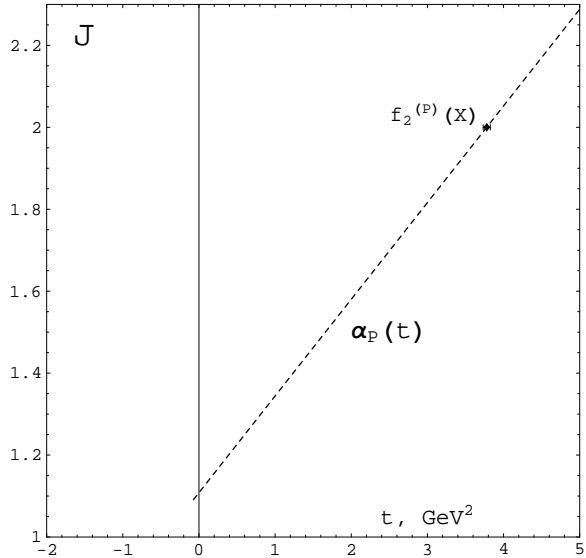}
\caption{Expected qualitative behavior of the real part of the Pomeron Regge trajectory in the resonance region.}
\label{pom2}
\end{center}
\end{figure}

In particular, some of the produced particles may be glueballs, {\it i.e.}, hadrons with prevailing gluon content, which have not been discovered yet. One of the most 
promising light tensor glueball candidates is the low-mass resonance of spin 2 related to the Regge trajectory (see Fig. \ref{pom2}) of the 
soft Pomeron (a Reggeon which dominates in the elastic scattering of protons at ultrahigh energies). In \cite{godizov2}, some partial widths of decay to pairs of light 
mesons were estimated for this resonance (further, we call it $f^{(\rm P)}_2(X)$) in the framework of the Regge-eikonal approach. However, for reliable identification of 
$f^{(\rm P)}_2(X)$ among other vacuum resonances produced exclusively at the RHIC or the LHC, we need both to know its branching ratios and to estimate the integrated and 
differential cross-sections of reaction $p+p\to p+f^{(\rm P)}_2(X)+p$.

The aim of this eprint is to provide such an estimation with the help of the simplest Regge-eikonal model applied earlier to the high-energy elastic scattering 
\cite{godizov} and single diffractive dissociation (SDD) \cite{godizov3} of protons.

\section*{2. The model}

In the region of high values of the collision energy and low values of the proton momentum transfers, the cross-section of exclusive diffractive production of glueball 
$f^{(\rm P)}_2(X)$ can be represented as 
\begin{equation}
\label{sig}
\sigma^{(\lambda)}_{p\,+\,p\,\to\,p\,+\,f^{(\rm P)}_2(X)\,+\,p}(\sqrt{s}) = \frac{1}{2s}\int
\frac{|T^{(\lambda)}|^2\,d^3p'_1\,d^3p'_2\,d^3k}{2p'^0_1(2\pi)^3\;2p'^0_2(2\pi)^3\;2k^0(2\pi)^3}\,(2\pi)^4\delta^4(p_1+p_2-p'_1-p'_2-k)\approx
$$
$$
\frac{1}{512\pi^4s}\int|T^{(\lambda)}(s,\xi_1,\xi_2,t_1,t_2,\phi)|^2\,\delta(\xi_1\xi_2s-(\vec\Delta_{1\perp}+\vec\Delta_{2\perp})^2-M_{f_2}^2)
\,d\xi_1\,d\xi_2\,dt_1\,dt_2\,d\phi\,,
\end{equation}
where $p_i$ and $p'_i$ are the 4-momenta of incoming and outgoing protons, $s = (p_1+p_2)^2$, vectors $\vec\Delta_{i\perp}$ are the tranverse components of 
$\Delta_i\equiv p_i-p'_i$ ($t_i=\Delta_i^2\approx -|\vec\Delta_{i\perp}|^2/(1-\xi_i)$), $k = \Delta_1 + \Delta_2$ is the 4-momentum of the produced tensor state, 
$\xi_i\ll 1$ are the energy fractions lost by the diffractively scattered protons, $\lambda$ is the produced particle helicity, $\phi$ is the angle between 
$\vec\Delta_{1\perp}$ and $\vec\Delta_{2\perp}$, $M_{f_2}$ is the produced resonance mass, and $T^{(\lambda)}$ is the full helicity amplitude of the 
$f^{(\rm P)}_2(X)$ CEP.

If the produced resonance has a significant decay width $\Gamma_{f_2}$, then the following replacement should be made in (\ref{sig}): 
$\delta(\xi_1\xi_2s-(\vec\Delta_{1\perp}+\vec\Delta_{2\perp})^2-M_{f_2}^2)\to\frac{1}{\pi}\frac{M_{f_2}\Gamma_{f_2}}
{(M_{f_2}^2+\,(\vec\Delta_{1\perp}+\vec\Delta_{2\perp})^2\,-\,\xi_1\xi_2s)^2\,+\,M_{f_2}^2\Gamma_{f_2}^2}\,$.

Next, constructing the double Pomeron fusion vertex in terms of independent tensor structures $\Delta_1^\mu\Delta_1^\nu$, $\Delta_1^\mu k^\nu$, $k^\mu\Delta_1^\nu$, 
$k^\mu k^\nu$, and $g^{\mu\nu}$, and taking account of the symmetry, transversality, and tracelessness of the produced glueball helicity states $e^{(\lambda)}_{\mu\nu}(k)$, 
we come to the expression for the bare helicity amplitudes of the $f^{(\rm P)}_2(X)$ exclusive production:
$$
T_{bare}^{(\lambda)}(s\,,\,\xi_1\,,\,\xi_2\,,\,\vec\Delta_{1\perp}\,,\,\vec\Delta_{2\perp}) = 
\left(i+\tan\frac{\pi(\alpha_{\rm P}(t_1)-1)}{2}\right)\left(i+\tan\frac{\pi(\alpha_{\rm P}(t_2)-1)}{2}\right)
\times
$$
\begin{equation}
\label{tripamp}
\times\;\left(\frac{1}{\xi_1}\right)^{\alpha_{\rm P}(t_1)}\left(\frac{1}{\xi_2}\right)^{\alpha_{\rm P}(t_2)}
\pi^2\alpha'_{\rm P}(t_1)\,\alpha'_{\rm P}(t_2)\;g_{pp\rm P}(t_1)\;g_{pp\rm P}(t_2)\;g_{\rm P\rm P f_2}(t_1\,,\,t_2)
\;\frac{\Delta_1^\mu\Delta_1^\nu}{s_0}\;e^{(\lambda)}_{\mu\nu}(k)\,,
\end{equation}
where $\alpha_{\rm P}(t)$ is the Regge trajectory of the Pomeron, $g_{pp\rm P}(t)$ is the Pomeron coupling to proton, $s_0=1$ GeV$^2$ is the unit of measurement, and 
$g_{\rm P\rm P f_2}(t_1,t_2)$ is the structure function related to the structure $\Delta_1^\mu\Delta_1^\nu$ in the double Pomeron fusion vertex. The factors 
$\pi\alpha'_{\rm P}$ are singled out within the Regge residue for the same reasons as in the cases of elastic scattering \cite{godizov} and high-missing-mass SDD 
\cite{godizov3}.

\begin{figure}[ht]
\epsfxsize=16.9cm\epsfysize=5.7cm\epsffile{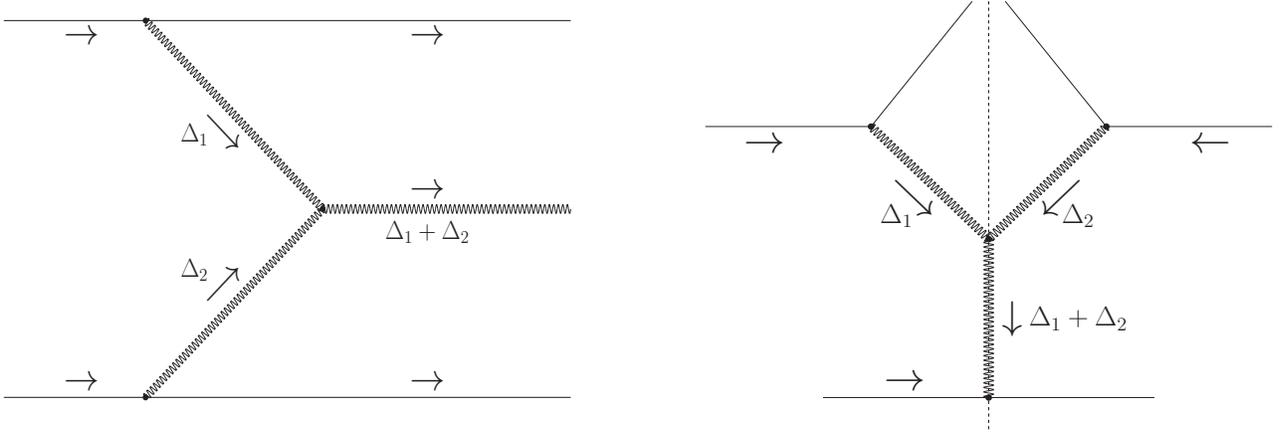}
\caption{The diagrams for the central exclusive production of $f^{(\rm P)}_2(X)$ via two-Pomeron fusion (left) and 
the single diffractive dissociation of proton at high missing masses (right).}
\label{diagr2}
\end{figure}

Comparing the diagram for the $f^{(\rm P)}_2(X)$ exclusive production (the left picture in Fig. \ref{diagr2}) with the triple-Pomeron diagram for the SDD 
of proton at high missing masses (the right picture in Fig. \ref{diagr2}), one immediately pays attention to some geometrical 
likeness between these two diagrams. Indeed, the vertex of two-Pomeron fusion to $f^{(\rm P)}_2(X)$ seems to be related to the triple-Pomeron vertex of SDD. 

To establish that relation between $g_{\rm P\rm P f_2}(t_1\,,\,t_2)$ and the triple-Pomeron vertex function $g_{3\rm P}(t_1,t_2,t_3)$ we need, first, to consider the 
expression for the SDD triple-Pomeron interaction amplitude (below we represent it in the form used in \cite{godizov3}), 
$$
T_{\rm 3P}(s\,,\,M_X\,,\,\vec\Delta_1\,,\,\vec\Delta_2) \approx \frac{1}{s}\left(i+\tan\frac{\pi(\alpha_{\rm P}(t_1)-1)}{2}\right)
\left(-i+\tan\frac{\pi(\alpha_{\rm P}(t_2)-1)}{2}\right)\times
$$
\begin{equation}
\times\;g_{pp\rm P}(t_1)\;g_{pp\rm P}(t_2)\;g_{pp\rm P}(t_3)\;g_{3\rm P}(t_1\,,\,t_2\,,\,t_3)\;\times
\label{tripleamp}
\end{equation}
$$
\times\;\pi^3\alpha'_{\rm P}(t_1)\,\alpha'_{\rm P}(t_2)\,\alpha'_{\rm P}(t_3)\;
\left(\frac{1}{\xi}\right)^{\alpha_{\rm P}(t_1)+\alpha_{\rm P}(t_2)}
\left(\frac{M_X^2}{2s_0}\right)^{\alpha_{\rm P}(t_3)}\,,
$$
where $M_X$ is the missing mass and $\xi=(M_X^2-m_p^2)/s$ is the energy fraction lost by the diffractively scattered proton, and, second, to replace the exchange by 
that Pomeron which carries 4-momentum $\Delta_1+\Delta_2$ ($(\Delta_1+\Delta_2)^2\equiv t_3$) by the exchange by that virtual particle $f_J$ of spin $J$ and mass $m_J$ 
which is related to the Pomeron Regge trajectory. Particularly, such a partial de-Reggeization implies the following replacements:
\begin{equation}
\alpha_{\rm P}(t_3)\to J\,,\;\;\;\;
\alpha'_{\rm P}(t_3)\to \frac{1}{m^2_J-t_3}\,,\;\;\;\;g_{pp\rm P}(t_3)\to g^{(J)}_{pp\rm P}(t_3)\,,
\label{deregge}
\end{equation}
$$
g_{3\rm P}(t_1\,,\,t_2\,,\,t_3)\to g^{(J)}_{\rm P\rm P f_J}(t_1\,,\,t_2\,,\,t_3)\,,
$$
where $g^{(J)}_{pp\rm P}(t_3)$ is the structure function related to the tensor structure ${p_2}^{\alpha_1}...{p_2}^{\alpha_J}$ in the tensor current of the proton 
carrying 4-momentum $p_2$ in the initial state, and $g^{(J)}_{3\rm P}(t_1\,,\,t_2\,,\,t_3)$ is the structure function related to the tensor structure 
${\Delta_1}^{\alpha_1}...{\Delta_1}^{\alpha_J}$ in the partially de-Reggeized triple-Pomeron vertex (these tensor structures dominate in the kinematic region 
$M_X\gg$ 1 GeV, because $M_X^2\approx 2(\Delta_1 p_2)$ in that range). 

Now it is obvious that $g_{\rm P\rm P f_2}(t_1\,,\,t_2)\equiv g^{(2)}_{\rm P\rm P f_2}(t_1,t_2,M_{f_2}^2)$, {\it i.e.}, it corresponds to the triple-Pomeron vertex 
function $g_{3\rm P}(t_1\,,\,t_2\,,\,t_3)$ in the limit $\{t_3\to k^2=M_{f_2}^2\,,\;\alpha(t_3)\to J=2\}$.

For quantitative predictions we need, first of all, to fix the model degrees of freedom, namely, the unknown functions $\alpha_{\rm P}(t)$, $g_{pp\rm P}(t)$, and 
$g_{\rm P\rm P f_2}(t_1\,,\,t_2)$. The Pomeron Regge trajectory and the Pomeron coupling to nucleon should be the same as in the $pp$ elastic scattering \cite{godizov}: 
\begin{equation}
\label{pomeron}
\alpha_{\rm P}(t) = 1+\frac{\alpha_{\rm P}(0)-1}{1-\frac{t}{\tau_a}}\,,\;\;\;\;g_{pp\rm P}(t)=\frac{g_{pp\rm P}(0)}{(1-a_gt)^2}\,,
\end{equation}
where the free parameters take on the values presented in Table \ref{tab1}.
\begin{table}[ht]
\begin{center}
\begin{tabular}{|l|l|}
\hline
\bf Parameter          & \bf Value                   \\
\hline
$\alpha_{\rm P}(0)-1$  & $0.109$            \\
$\tau_a$               & $0.535$ GeV$^2$  \\
$g_{pp\rm P}(0)$       & $13.8$ GeV         \\
$a_g$                  & $0.23$ GeV$^{-2}$ \\
\hline
\end{tabular}
\end{center}
\vskip -0.2cm
\caption{The parameter values for (\ref{pomeron}) obtained via fitting to the elastic scattering data.}
\label{tab1}
\end{table}

As well, it was argued in \cite{godizov3} that 
\begin{equation}
g_{3\rm P}(t_1\,,\,t_2\,,\,t_3)\approx g_{3\rm P}(0\,,\,0\,,\,0)\approx 0.64\;{\rm GeV}
\label{sing}
\end{equation}
in the kinematic range relevant for SDD. The main hypothesis we use further is the assumption that this equality may be extended to the $f^{(\rm P)}_2(X)$ CEP region:
\begin{equation}
g_{\rm P\rm P f_2}(t_1\,,\,t_2)\equiv g_{3\rm P}(t_1\,,\,t_2\,,\,M_{f_2}^2)\approx g_{3\rm P}(t_1\,,\,t_2\,,\,0)\approx g_{3\rm P}(0\,,\,0\,,\,0)\,.
\label{vertex}
\end{equation}

Having fitted $\alpha_{\rm P}(t)$ and $g_{pp\rm P}(t)$ to the elastic scattering data and the value of $g_{3\rm P}(0\,,\,0\,,\,0)$ to the data on the proton SDD, 
and, next, having made assumption (\ref{vertex}), we are able to estimate the bare amplitude (\ref{tripamp}) of the $f^{(\rm P)}_2(X)$ CEP.  

To calculate the corresponding CEP cross-section (\ref{sig}) we need to take account of the multi-Pomeron exchanges between the incoming protons and the outgoing 
ones.\footnote{For detailed discussion of the importance and significance of such absorptive corrections in high-energy CEP, see papers \cite{ryutin} and \cite{khoze} and 
references therein.} It can be done in the same way as for the proton SDD cross-section \cite{godizov3}. Then, the full helicity amplitude can be approximated by the 
expression \begin{equation}
T^{(\lambda)}(s,\xi_1,\xi_2,t_1,t_2,\phi) \approx T_{bare}^{(\lambda)}(s,\xi_1,\xi_2,\vec\Delta_{1\perp},\vec\Delta_{2\perp})\,+
$$
$$
+\,\frac{1}{16\pi^2s}\int d^2\vec q_{1\perp}\;A(s,-\vec q_{1\perp}^{\;2})\,
T_{bare}^{(\lambda)}(s,\xi_1,\xi_2,\vec\Delta_{1\perp}-\vec q_{1\perp},\vec\Delta_{2\perp}+\vec q_{1\perp})\,+
\end{equation}
$$
+\,\frac{1}{16\pi^2s}\int d^2\vec q_{2\perp}\;T_{bare}^{(\lambda)}(s,\xi_1,\xi_2,\vec\Delta_{1\perp}-\vec q_{2\perp},\vec\Delta_{2\perp}+\vec q_{2\perp})\,
A(s,-\vec q_{2\perp}^{\;2})\,+
$$
$$
+\,\frac{1}{(16\pi^2s)^2}\int d^2\vec q_{1\perp} d^2\vec q_{2\perp}\;A(s,-\vec q_{1\perp}^{\;2})\,
T_{bare}^{(\lambda)}(s,\xi_1,\xi_2,\vec\Delta_{1\perp}-\vec q_{1\perp}-\vec q_{2\perp},\vec\Delta_{2\perp}+\vec q_{1\perp}+\vec q_{2\perp})\,A(s,-\vec q_{2\perp}^{\;2})\,,
\label{full}
$$
where the absorption subamplitudes $A(s,t)$ are computed in the following way:
$$
A(s,t) = 4\pi s\int_0^{\infty}db^2\,J_0(b\sqrt{-t})\,(e^{i\delta(s,b)}-1)\,,
$$
\begin{equation}
\delta(s,b) = \frac{1}{16\pi s}\int_0^{\infty}d(-t)\,J_0(b\sqrt{-t})\,\delta_{\rm P}(s,t)\,,
\label{eikrepr}
\end{equation}
$$
\delta_{\rm P}(s,t) = \left(i+\tan\frac{\pi(\alpha_{\rm P}(t)-1)}{2}\right)g^2_{pp\rm P}(t)\;\pi\alpha'_{\rm P}(t)\left(\frac{s}{2s_0}\right)^{\alpha_{\rm P}(t)}\,.
$$

\section*{3. Application to the WA102 data and predictions for the CEP of $f^{(\rm P)}_2(X)$ at the LHC}

It was shown in \cite{godizov2} that $f_2(1950)$ is the most promising candidate for the status of the light tensor glueball lying on the Pomeron Regge trajectory. 
If we put $M_{f_2}=$ 1.944 GeV and\linebreak $\Gamma_{f_2}=$ 472 MeV \cite{pdg}, then we obtain the following prediction for the $f_2(1950)$ CEP cross-section 
at $\sqrt{s}=29.1$ GeV integrated over the range \{$M_{f_2}-\Gamma_{f_2}<\sqrt{k^2}<M_{f_2}+\Gamma_{f_2}$ and\linebreak $\sqrt{|t_{1,2}|}<1$ GeV\}:
\begin{equation}
\sigma^{model}_{p\,+\,p\,\to \,p\,+\,f_2(1950)\,+\,p}(29.1\,{\rm GeV})=
\sum_\lambda\sigma^{(\lambda)}_{p\,+\,p\,\to \,p\,+\,f_2(1950)\,+\,p}(29.1\,{\rm GeV})\approx 0.33\,\mu b\,.
\label{sig2}
\end{equation}
This estimation is 8.5 times less than the measured experimental value \cite{kirk}: 
\begin{equation}
\sigma^{WA102}_{p\,+\,p\,\to \,p\,+\,f_2(1950)\,+\,p}(29.1\,{\rm GeV})= (2.788\pm 0.175)\,\mu b\,.
\label{sig3}
\end{equation}

Such a divergence is due to the fact that the characteristic value $\xi_{1,2}\sim 0.07$ of the energy fractions lost by protons in the WA102 experiment
is so high that the combined contribution of the secondary Reggeon exchanges to the bare amplitude of CEP may be comparable to the double Pomeron exchange term 
(\ref{tripamp}) or even may dominate over it. Therefore, the obtained model underestimation of the $f_2(1950)$ production cross-section seems quite natural. 

\begin{figure}[ht]
\vskip -0.5cm
\epsfxsize=8.2cm\epsfysize=8.2cm\epsffile{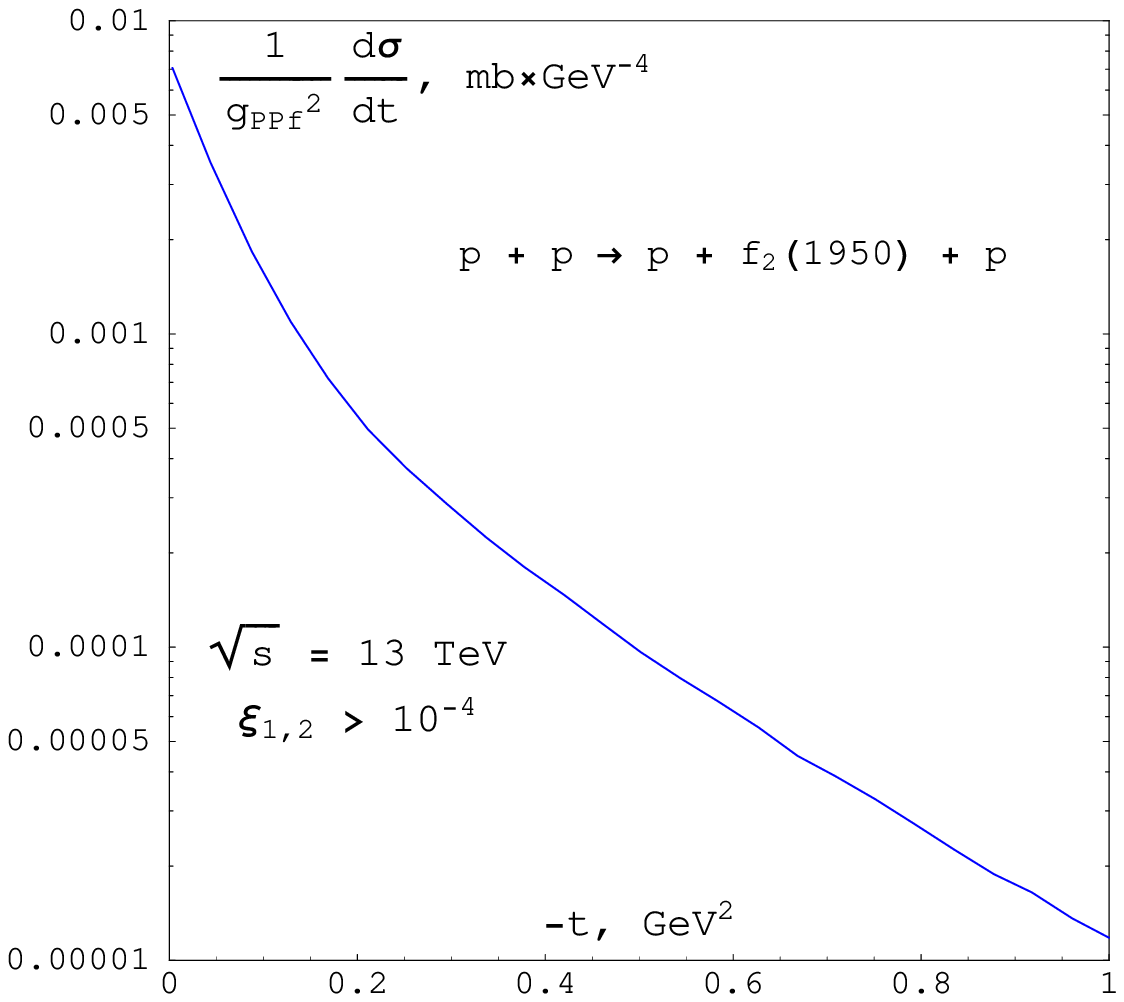}
\vskip -8.1cm
\hskip 8.95cm
\epsfxsize=8.0cm\epsfysize=8.0cm\epsffile{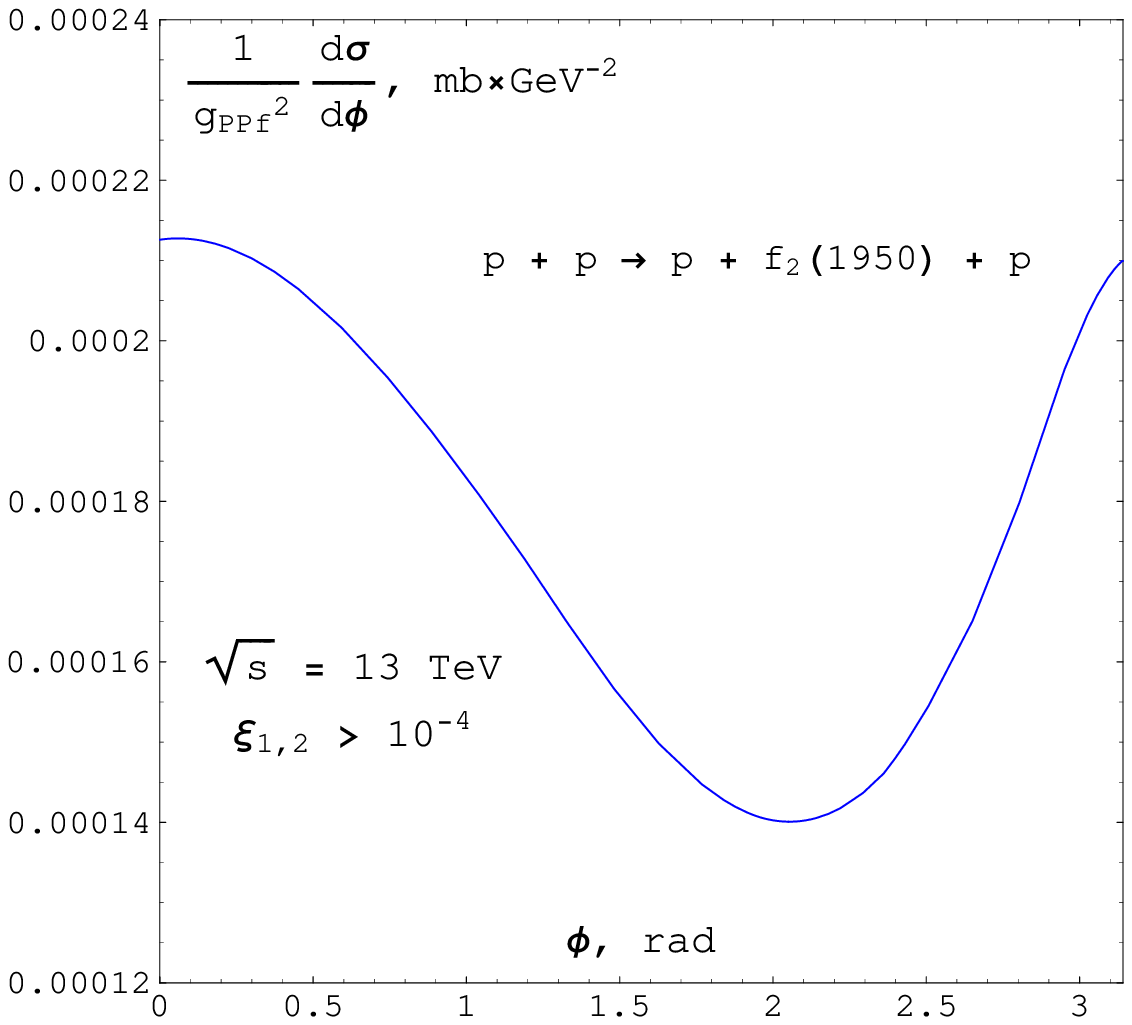}
\caption{The model $t$- and $\phi$-distributions of the $f_2(1950)$ CEP at $\sqrt{s}=$ 13 TeV obtained via integration over the range 
\{$M_{f_2}-\Gamma_{f_2}<\sqrt{k^2}<M_{f_2}+\Gamma_{f_2}$ and $\xi_{1,2}>10^{-4}$\}.}
\label{diff}
\end{figure}

The characteristic value of $\xi_{1,2}$ at the LHC is much lower and the dominance of the two-Pomeron fusion term in the bare amplitude is guaranteed. The 
model computation of the $f_2(1950)$ CEP cross-section at $\sqrt{s}=$ 13 TeV integrated over the kinematic range 
\{$\xi_{1,2}>10^{-4}$,\linebreak $\sqrt{|t_{1,2}|}<1$ GeV, and $M_{f_2}-\Gamma_{f_2}<\sqrt{k^2}<M_{f_2}+\Gamma_{f_2}$\} yields the following value:
\begin{equation}
\sigma^{(\xi_{1,2}>10^{-4})}_{p\,+\,p\,\to \,p\,+\,f_2(1950)\,+\,p}(13\,{\rm TeV})\approx 0.22\,\mu b\,.
\label{sig4}
\end{equation}
The corresponding model $t$- and $\phi$-distributions are presented in Fig. \ref{diff}.

\section*{4. Discussion}

The above-considered model is grounded on the fact that the vertex function $g_{\rm P\rm P f_2}(t_1\,,\,t_2)$ related to the CEP of the discussed light tensor 
glueball $f^{(\rm P)}_2(X)$ via two-Pomeron fusion and the triple-Pomeron vertex function $g_{3\rm P}(t_1\,,\,t_2\,,\,t_3)$ which governs the SDD of proton at 
high energies and high missing masses are just different branches of the same analytic function. This fact is model-independent.

As well, for calculation of the corresponding CEP cross-sections we assumed the negligibility of the nontrivial analytic structure of function 
$g_{3\rm P}(x\,,\,y\,,\,z)$ in the unique kinematic region which covers both the ranges relevant for the reaction $p+p\to p+f^{(\rm P)}_2(X)+p$ and the 
high-missing-mass SDD of proton. Having estimated the cross-section of the $f^{(\rm P)}_2(X)$ CEP and its partial widths of decay 
to pairs of light mesons (see paper \cite{godizov2}), we can try to distinguish this tensor glueball among other vacuum resonances produced exclusively at the 
RHIC or the LHC and decaying through, say, the $\pi^+\pi^-$ and $K^+K^-$ channels.\footnote{In the author's opinion, the most promising candidate is $f_2(1950)$.}

The weakest point of the proposed model is assumption (\ref{vertex}) which is an extension of assumption (\ref{sing}) and seems to be very strong. In its turn, 
assumption (\ref{sing}) is based just on the fact that it allows to obtain an estimation of the SDD cross-section logarithmic $t$-slope in the range 
0.05 GeV$^2\le -t\le 0.11$ GeV$^2$ at $\sqrt{s}=1.8$ TeV, $B^{model}=11.7$ GeV$^{-2}$ \cite{godizov3}, which is consistent\linebreak with the value measured by 
the E-710 Collaboration, $B^{E-710}=(10.5\pm 1.8)$ GeV$^{-2}$ \cite{E710}. Other data on the $t$-behavior of the proton SDD cross-section are available only 
in those kinematic ranges wherein the secondary Reggeon exchange contributions are comparable to the triple-Pomeron term (for details, see 
\cite{godizov3}). The CMS data on the SDD $\xi$-distribution at $\sqrt{s}=7$ TeV and $10^{-5.5}<\xi<10^{-2.5}$ \cite{cms} do not allow to confirm 
the adequacy of assumption (\ref{sing}) in the relevant kinematic range. The replacement, say, 
$g_{3\rm P}(0\,,\,0\,,\,0)\to g_{3\rm P}(0\,,\,0\,,\,0)\cdot e^{a(t_1+t_2+t_3)}$, where\linebreak $a=0.5$ GeV$^{-2}$ yields just a slight (5--7\%) decrease of 
$d\sigma^{SDD}/d\xi$ in the above-mentioned\linebreak $\xi$-interval, which can be easily compensated by an increase of $g_{3\rm P}(0\,,\,0\,,\,0)$. Thus, at first 
glance, approximation (\ref{sing}) has a very weak phenomenological foundation even at low negative values of $t_1$, $t_2$, and $t_3$, and, hence, its extension up to 
$t_3\to M_{f_2}^2\sim 4$ GeV$^2$ seems unjustified. 

However, from the phenomenological standpoint, assumptions (\ref{sing}) and (\ref{vertex}) are supported by the established behavior of the Pomeron couplings to 
various light mesons and photons. It was shown in \cite{godizov2} that approximations $g_{hh\rm P}(t)\approx g_{hh\rm P}(0)$ (here 
$h=\pi\,,\,K\,,\,\rho\,,\,\omega\,,\,\phi\,,\,\gamma$) are not only consistent with available data on the corresponding exclusive reactions of 
meson-proton and photon-proton scattering, but, being extended to the interval $0<t<M_{f_2}^2$, allow to obtain the estimations of the quantities 
$\Gamma_{f_2(1950)\to\gamma\gamma}\,\Gamma_{f_2(1950)\to K^+K^-}/\Gamma_{f_2(1950)\to X}$ and 
$\Gamma_{f_2(1950)\to\gamma\gamma}\,\Gamma_{f_2(1950)\to \pi^0\pi^0}/\Gamma_{f_2(1950)\to X}$, which are quite in agreement with the 
Belle Collaboration data \cite{kaon,pion}. Hence, we could expect that the function $g_{3\rm P}(t_1\,,\,t_2\,,\,t_3)$ 
(the Pomeron coupling to the Pomeron) in the region \{$-1$ GeV$^2<t_3<M_{f_2}^2$; $-1$ GeV$^2<t_{1,2}<0$\} is not an exception to the rule, and, 
thus, approximation (\ref{vertex}) should be kept in mind as a quite probable property of the triple-Pomeron coupling.

Certainly, it is possible that the true value of $g_{3\rm P}(0\,,\,0\,,\,M_{f_2}^2)$ is not equal to the value of $g_{3\rm P}(0\,,\,0\,,\,0)$. 
However, in view of the aforesaid, we could expect, at least, the validity of a much weaker assumption in the range of low negative $t_{1,2}$: 
\begin{equation}
g_{\rm P\rm P f_2}(t_1\,,\,t_2)\approx g_{\rm P\rm P f_2}(0\,,\,0)\equiv g_{3\rm P}(0\,,\,0\,,\,M_{f_2}^2)\,.
\label{vertex2}
\end{equation}
In this case, we can not provide a prediction for the $f^{(\rm P)}_2(X)$ CEP cross-section, but the distributions 
presented in Fig. \ref{diff} remain unchanged.

Moreover, relations analogous to (\ref{vertex2}) could take place in the region of low $|t_{1,2}|$ for the CEP of various $C$-even isoscalar vacuum mesons. Such a variant 
does not seem exotic, if we consider the $t$-evolution of the $\rho$-Reggeon and $f$-Reggeon couplings to pion in the region $t>0$. Application of formula (A.4) from the 
Appendix of \cite{godizov2} to the $\pi\pi$ decays of mesons $\rho(770)$, $\rho_3(1690)$, $f_2(1270)$, and $f_4(2050)$ \cite{pdg} yields the 
following ratios:
\begin{equation}
\frac{|g_{\pi\pi\rho}(M_{\rho_3(1690)}^2)|}{|g_{\pi\pi\rho}(M_{\rho(770)}^2)|}=1.1\pm 0.05\,,\;\;\;\;
\frac{|g_{\pi\pi f}(M_{f_4(2050)}^2)|}{|g_{\pi\pi f}(M_{f_2(1270)}^2)|}=0.56\pm 0.06\,.
\label{rats}
\end{equation}
Thus, the couplings of these Reggeons to pion have very weak $t$-dependence in the resonance region. Consequently, we could expect that the dependence of the 
Pomeron-Pomeron-meson vertex functions (the meson couplings to the Pomeron) on $t_1$ and $t_2$ in the range $|t_{1,2}|<1$ GeV$^2$ might be weak, at 
least, for some of vacuum mesons.

Then, the distributions over the kinematic ranges relevant for the nucleon-nucleon elastic scattering, SDD of proton, and CEP of some of vacuum resonances 
(including $f^{(\rm P)}_2(X)$) turn out strongly correlated, since they are determined by the $t$-behavior of $\alpha_{\rm P}(t)$ and $g_{pp\rm P}(t)$ only, while the 
corresponding Pomeron-Pomeron-meson vertex functions can be treated just as some constants which determine the values of the meson CEP cross-sections, but 
have no influence on the form of the distributions over kinematic variables. Therefore, the proposed approach could be very useful in phenomenology 
of high-energy CEP of light vacuum mesons (see \cite{ryutin2} for more details).

In the very end, it should be pointed out that approximations (\ref{vertex}) and (\ref{vertex2}) are still just an assumptions, though they have reasonable phenomenological 
grounds. Besides, these restrictions are so stiff that they can be easily confirmed or discriminated by the forthcoming experimental data on 
the CEP of light vacuum resonances from the RHIC and the LHC.

\subsection*{Acknowledgements} This work was reported at the DTP-IHEP seminar on 04.09.2018. The author is indebted to his colleagues from the IHEP Division of 
Theoretical Physics for their questions and useful criticism.

\end{document}